# The Global Care Ecosystems of 3D Printed Assistive Devices

SAIPH SAVAGE
Northeastern University, s.savage@northeastern.edu

CLAUDIA FLORES-SAVIAGA
Northeastern University, floressaviaga.c@northeastern.edu

RACHEL RODNEY
University of Washington

LILIANA SAVAGE
Universidad Autonoma de Mexico (UNAM)

JON SCHULL
Enable.org

JENNIFER MANKOFF
University of Washington

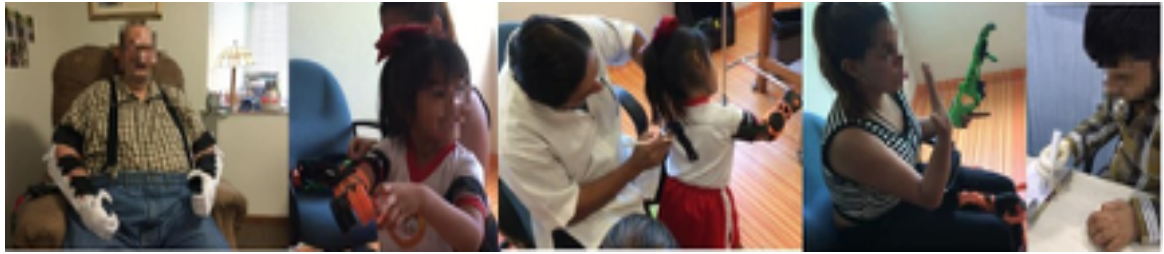

**Figure 1:** We interviewed multiple stakeholders from different countries in the care ecosystems that have emerged around recipients and their 3D printed assistive devices for the upper extremities. The figure shows some of the recipients, clinicians, and makers from different parts of the world who took part in our research.

The popularity of 3D printed assistive technology has led to the emergence of new ecosystems of care, where multiple stakeholders (makers, clinicians, and recipients with disabilities) work toward creating new upper limb prosthetic devices. However, despite the increasing growth, we currently know little about the differences between these care ecosystems. Medical regulations and the prevailing culture have greatly impacted how ecosystems are structured and stakeholders work together, including whether clinicians and makers collaborate. To better understand these care ecosystems, we interviewed a range of stakeholders from multiple countries, including Brazil, Chile, Costa Rica, France, India, Mexico, and the U.S. Our broad analysis allowed us to uncover different working examples of how multiple stakeholders collaborate within these care ecosystems and the main challenges they face. Through our study, we were able to uncover that the ecosystems with multi-stakeholder collaborations exist (something prior work had not seen), and these ecosystems showed increased success and impact. We also identified some of the key follow-up practices to reduce device abandonment. Of particular importance are to have ecosystems put in place follow up practices that integrate formal agreements and compensations for participation (which do not need to be just monetary). We identified that these features helped to ensure multi-stakeholder involvement and ecosystem sustainability. We finished the paper with socio-technical recommendations to create vibrant care ecosystems that include multiple stakeholders in the production of 3D printed assistive devices.

# 1 INTRODUCTION

In recent years, 3D printing has become extremely popular. One common use is assistive technology (AT) [52, 133], including prosthetics for people with upper extremity limb loss [136, 140, 141]. This group includes almost two million people in the United States alone [6]. In the design space of 3D printed AT for upper extremity limb loss, many 3D printed devices resemble arms, hands, or fingers [28, 140, 141]. The popularity of such technology is understandable since 3D printing facilitates the creation of customized assistive devices, which allow people with upper extremity limb loss to have AT tailored to their desired activities [69]. For example, a number of custom-built 3D printed designs have emerged to enable people with upper limb loss to easily play a musical instrument or ride a bike [42].

The popularity of this type of AT has also led to the emergence of care ecosystems that focus on supporting the needs of people with upper limb loss who want to use such devices [42, 51, 53, 73, 119]. Perhaps the most famous and widespread is e-NABLE and its international network of community chapters [120].

These ecosystems of care can include multiple stakeholders: (a) medical professionals (clinicians), (b) technologists/manufacturers (makers) who can 3D print AT, and (c) people with an upper limb difference (recipients) who use the 3D printed AT. However, despite the popularity and potential of such ecosystems, there is limited information available about the type of collaboration structures these ecosystems enjoy. Most prior work primarily studied individual stakeholders [51, 59, 100]. As a result, we have a limited understanding of how these ecosystems function.

Additionally, most previous work focused on a single country: the United States (e.g., [42, 51, 53, 73, 119]). This implicitly limits the range of ecosystem structures we can observe and study. For example, past work identified critical and practical limitations to the collaboration of clinicians and makers in the U.S. [48, 100]. This has unfortunately led to the understudy of positive collaboration examples. Consequently, we lack information about practices that might facilitate collective work between stakeholders. Positive multi-stakeholder collaborations only began to arise in the U.S. when the onset of COVID-19 caused structural barriers to collaborations to rapidly dismantle [50].

The lack of studies on the perspectives of all stakeholders means we cannot know the details of how such collaborations emerged nor the challenges and opportunities the collaborations encountered. To address this gap, we drew our sample from multiple countries, including those where clinicians, makers, and recipients collaborate closely. We interviewed stakeholders based in seven different countries (Brazil, Chile, Costa Rica, France, India, Mexico, and the U.S.). Our work uncovered multiple examples of multi-stakeholder collaborations operating successfully outside of the U.S. context, which we attribute to the cultural and societal constraints that facilitate or hinder certain collaborations [128]. Our participants had different and even multiple roles within their ecosystems as either recipients, clinicians, and makers of 3D printed AT.

Our novel findings on ecosystem features include:

- The relationships between clinical involvement, successful follow-up, and increased safety in 3D printed assistive technology.
- The impact of formal agreements on sustained participation and follow up.
- The impact of funding on sustained recipient participation in low-resource regions.

To our knowledge, prior work (as summarized in Table 1) has not studied these ecosystem features.

**SUMMARY OF PAPER.** To summarize, our study explores variants of care ecosystems and what stakeholders of these ecosystems believe enables them to function. Based on our findings, we provide a rich overview of the different ways care ecosystems operate, the opportunities and challenges they face, and offer design recommendations for encouraging successful multi-stakeholder collaboration. Note that our goal was not to characterize all ecosystems that exist; we used examples of differences in how ecosystems function to better characterize the key features of care ecosystems.

## 2 RELATED WORK AND BACKGROUND ON GLOBAL HEALTHCARE SYSTEMS

Assistive Technology (AT) is defined as any piece of equipment used to maintain or improve the functional capabilities of individuals with disabilities [41]. Unfortunately, most AT is expensive and has difficulty adapting to recipients' changing needs [23, 30]. As a result, a great majority of AT is eventually abandoned [104]. In this literature review, we explored the intersection of AT and fabrication. We also aimed to provide background information about the healthcare systems of the multiple global stakeholders involved in these care ecosystems. Within this intersection, we first reviewed what is known about fabrication in clinical and maker communities, including the impact of 3D printed AT on recipients. We then strove to explore a small body of literature on ecosystems that fabricate AT. We summarized the most relevant literature on AT and fabrication in Table 1. Other than one study with interviews in the U.S and Canada [73], and a very recent study that includes some "global" communities during the COVID-19 pandemic [50], all the studies we could find looked only at U.S. populations.

### 2.1 Clinicians and 3D Printed Assistive Devices

When 3D printing began to catch the attention of clinicians in the 1990s [112], it was viewed as a possible way to bring down costs. When 3D printing is used directly within clinical settings, it is typically referred to as "medical making" [2, 9, 48, 73]. Clinicians in this context have to balance incorporating innovation in their jobs and being explicitly committed to the oath of "do not harm" [2, 5, 73, 105, 116].

Clinicians have expressed concern that 3D printed assistive devices could harm the populations they support [48]. A low-quality device may threaten not only the recipient's ability to use their limb, but also threaten their life [14, 68]. Clinicians are therefore careful to understand how devices can potentially injure recipients. Studies in the U.S. have found that clinicians have hesitated to recommend 3D printing technology to patients (e.g., [74, 91]). Within the U.S. context, pay structures and job insecurity may also induce clinicians not to encourage the use of 3D printing technologies [67].

Additionally, a lack of training in fabrication, design, and 3D printing, in general, can also deter clinicians from considering 3D printed devices as an option in the clinical setting [5, 91]. However, with the onset of COVID-19, medical making increased significantly in the clinical context [72]. Nonetheless, it was challenging to go beyond prototyping [71]. Regulatory and cultural issues also impacted how maker communities engaged in such efforts [50].

| | | | | | | |
|---|---|---|---|---|---|---|
| [48] | e-NABLE | Experience report | 1 | 5 | 0 | U.S. |
| [73] | Clinic | Interviews | 10 | 3 | 0 | U.S./Canada |
| [51] | Clinic | Case Study | 4 | 0 | 0 | U.S. |
| [74] | Clinic | Interviews | 10 | 3 | 0 | U.S. |
| [50] | Clinic | Participant Observation Study | 0 | 14 | 0 | Global (10/14 U.S.) |
| [72] | Clinic | Interviews | 13 | 0 | 0 | U.S. |
| [5] | Clinic | Interviews | 8 | 0 | 0 | U.S. |
| [2] | Clinic | Interviews | 10 | 0 | 0 | U.S. |
| [100] | e-NABLE | Interviews | 3 | 11 | 0 | U.S. |

Savage et al.| Ref | Focus | Method | C | M | R | Country |
|---|---|---|---|---|---|---|
| [46] | DIY Recipient-Maker in e-NABLE | Auto-ethnography | 0 | 1 † | | U.S. |
| [99] | DIY Makers | Interviews | 0 | 4 | 0 | U.S. |
| [1] | DIY Makers | Experience Report | 4 | 4 | 1 | U.S. |
| [57] | DIY Recipient-Maker | Interviews | 0 | 10 | 1 | U.S. |
| [63] | Co-Creation with Recipient Makers | Experience Report | 4 | 9 | 13 | U.S. |
| [17] | e-NABLE | Interviews | 0 | 0 | 14 | U.S. |
| [*] | Ecosystem Understanding (e-Nable; Companies; Clinics) | Interviews | 6 | 9 | 16 | 7 Countries |

**Table 1:** Overview of prior work. For each prior research paper, we present: Focus (which includes: e-Nable; Clinic (which includes medical makers and/or clinicians associated with a clinic or hospital); DIY Makers not associated with e-NABLE; and DIY Recipient-Makers (people who are both recipients and DIY makers)); Research Method; number of participants in the prior work study who were either Clinicians; Makers, or Recipients (columns "C", "M", and "R"); and the Country from which participants were recruited. [*] refers to this paper. † indicates that the same participant fell into multiple categories.

## 2.2 Makers and 3D Printed Assistive Devices

In parallel to the increasing numbers of clinicians working with 3D printed AT, the global rise of do-it-yourself, non-medical (DIY) maker communities began to develop AT in the 2010s with the goal of bringing down costs (e.g., [52]). One of the most widespread examples is the design, fabrication, and distribution of 3D-printed assistive hands [21, 40, 43], pioneered by the e-NABLE community[1] [120]. The maker movement afforded an opportunity to disseminate low-cost solutions for healthcare and improve the design of health innovations through open-source collaborations [12, 52].

For example, the e-NABLE community, one of the largest of such organizations, coordinates the design, fabrication, and distribution of 3D printed AT, primarily devices shaped like hands. The e-NABLE community currently has over 7,000 members in over 45 countries. The majority of these maker communities originated through informal collaborations [54, 129]. They focus on populations with upper-limb differences for whom traditional AT is too expensive [12, 92]. The U.S. has several maker communities committed to a similar mission (e.g., the Open Hand Project, Open Bionics, e-NABLE, Limbitless) [120]. Many chapters and spin-offs of these communities have also emerged worldwide (see the map of e-NABLE chapters across countries[2]). Several private companies also use 3D printing to fabricate assistive technology [8, 25, 86, 127].

Many of these companies have adopted 3D printing to provide customers (recipients) with a higher degree of individual customization than what traditional technology offers [81]. Several were originally 3D printing manufacturing companies that partnered with social-service organizations to start providing low-cost custom assistive devices for people with disabilities (while making a small profit) [19]. Other firms arose from private companies with vast experience in building assistive technology, using traditional materials [76, 122]. These firms started to adopt 3D printing technology due to the customization it provides to their designs, while also being lower in cost [22, 76, 102]. Companies, in general, came to see this combination as a valuable business opportunity [64]. Several have already enjoyed long-established collaborations with hospitals [80, 102]. This likely facilitated mainstreaming the technology with recipients, while also including clinicians in the process.

---
[1] http://enablingthefuture.org/
[2] http://enablingthefuture.org/e-nable-community-chapters/

Similar to clinicians, makers are also concerned with the dangers of hurting recipients through their 3D printed innovations [58, 116]. However, studies in the U.S. suggest that makers may put more emphasis on prioritizing values, such as "being innovative" and "helping where you can" over "doing no harm" [58, 100, 138]. Consequently, makers may not constrain their behavior as much as clinicians [95, 106, 143]. Makers' lack of training in clinical work can also limit their awareness of risk [48]. Overall, research has argued that the devices produced by traditional maker communities in the U.S. may be problematic because they typically lack close connections with clinicians [48], a structural situation that has raised concerns about the health and safety implications of the ideals driving the work of these communities [16, 38, 45, 48]. However, to understand this phenomena more deeply, it is also important to note that on the other side of this, makers have argued that clinicians exaggerate the risks involved with 3D printed AT in order to gatekeep their field [144, 145]. Makers in these cases have contended that the psychological and social benefits of gifting recipients 3D printed AT overcomes the minor harms that the devices could inflict on recipients [23, 48, 146].

On the other hand, medical makers (i.e., clinicians who actively participate in the fabrication of 3D printed AT) have appeared to address the important limitations that traditional maker communities have encountered [48]. However, medical makers have also struggled to engage and collaborate with other stakeholders [50], which can impact the overall effectiveness and reach of the 3D printed AT that they fabricate.

## 2.3 Recipients of 3D Printed Assistive Technology

Recent research has started to look at how we can integrate the experiences and perspectives of the users of AT (recipients) to drive better designs [44, 63, 79]. Bennett et al. [17] interviewed adult U.S. recipients. They began to map the role that identity plays in the appropriation of 3D printed assistive devices and provided a much deeper picture of how 3D printing is changing what we understand assistive technology to be.

The work also provided design recommendations that integrate the perspectives and input of recipients. Similarly, other related research began reporting on the experience of integrating recipients into the making of AT [49, 57]. Studies have found that recipients have an interest in being involved in the process of creating and defining their own AT. However, these studies were again limited to a single region (Canada and the U.S.) This consequently limited the range of experiences studied to only represent people from that particular cultural context. Studies on recipients have other limitations. Hawthorn et al. [46] focused primarily on an interview with a single U.S. e-NABLE recipient. They note that we lack an understanding of the type of follow-up recipients are receiving. (Notice that similar to prior work, we define follow-up as the contact that the recipient receives after the device has been delivered). Another gap is understanding whether there is real value for recipients in the AT produced [103].

## 2.4 Ecosystems of Care

Care ecosystems are dynamic networks that emerge through connections between different stakeholders all working toward improving the care provided to a particular population [10, 125]. Following ecology terminology [31, 40, 94], we view a care ecosystem as interconnected groups of makers, clinicians, and recipients. We used the HCI literature on care ecosystems to characterize and study these networks [33]. The vast majority of prior research has focused on studying individual stakeholders instead of considering how different stakeholders operate together within the ecosystem or the type of work dynamics they have in place [36, 87, 89]. The limited number of studies of care ecosystems in 3D printed AT has affected the type of technologies developed. Prior work has tended to focus on individuals instead of ecosystems [24, 61, 132]. Only recently have we started to see the emergence of research focused on studying more than individual stakeholders [1, 50, 63, 72, 74, 85]., e.g. research focused on conducting workshops to identify the ways in which the stakeholders within the workshop could work together to co-design upper limb prostheses [1, 63].

Jones et al. [63] found that to facilitate co-designs with multiple stakeholders, it was key to provide a transparent information flow between the stakeholders. Similarly, Aflatoony et al. [1] identified that multi-stakeholder co-designs facilitated working within a synergistic manner, where the combined expertise was greater than the sum of the isolated skills of the different stakeholders. We now build off this prior research to study in the wild how people are managing to drive co-designs of 3D printed AT with multiple stakeholders. We are interested in identifying the type of strategies that have been put in place to enable such collaborations. Recent related research has made great strides in starting to study how medical makers in the U.S. collaborate with other stakeholders in the wild [1, 9, 50, 72]. While the collaborations studied in prior work are not focused on building 3D printed prostheses, they can help us further understand multi-stakeholder collaborations around DIY medical devices. Hofmann et al. [50] conducted an observational study on online forums and sites, attempting to address COVID-19 related needs and uncovered how medical makers have aimed to connect with other stakeholders to conduct collective action to fabricate medical devices. These sites are primarily U.S. based, and their aim is not always to produce DIY devices, but rather to have key players enable mass production for clinical settings. The research found that despite small victories, the stakeholders were hampered by a lack of connections, for example, with policy makers. Similarly, Lakshmi et al. [72] interviewed 13 U.S.-based medical makers to understand how during the COVID-19 pandemic, they collaborated with other stakeholders. The research revealed that to facilitate collaborations, medical makers had to become links between different institutions, maker communities, and wider regional industry networks. This enabled the production of medical devices that followed certain regulatory norms, while also considering particular human constraints. The work also identified several limitations of makers when connecting and working with different stakeholders. We use the findings of these studies to guide us in studying multi-stakeholder collaborations around 3D printed AT.

**SUMMARY OF RELATED WORK.** In summary, prior research has explored the potential of 3D printed AT from multiple perspectives, but it has:

- Primarily been limited to the U.S. context
- Rarely included recipients
- Rarely looked at larger care ecosystems

To address these gaps, our research broadened the stakeholders and regions studied, providing a more extensive lens to view the problems these care ecosystems face.

Given that our research aims to understand how multiple stakeholders from different parts of the world operate, we provide background information about the healthcare systems associated with the countries represented in our study in the following section.

## 2.5 Contextualizing Global Healthcare Systems

Although global healthcare systems are not easily summarized due to their variety and the different factors involved [121], we now provide a brief overview of some of the differences and similarities of the healthcare systems in the countries we studied. This will help contextualize our results.

*UNITED STATES.* The United States spends around 18% of its GDP on its health system, the largest percentage among all countries in the world [35]. Nevertheless, almost 28 million people in the U.S. have declared a lack of coverage for their medical expenses. The domestic healthcare system ranks 37 out of 191 countries, according to its performance [11, 142] (usually measured in terms of life expectancy, infant mortality, healthcare inequality, and other healthcare measures [7, 78]). While the U.S. healthcare system offers advanced technologies and is known to excel in the "doctor-patient relationship" [29], the healthcare model works primarily as a for-profit operation with limited access, large variations in equity based on income, and poor affordability for the average person [29].

Medical insurance is a critical element in access to healthcare services. Of note, 34% of patients in the United States are covered by private insurance while 37% enroll in national health insurance programs, such as Medicare or Medicaid to cover healthcare costs [35]. Finally, the U.S. healthcare system is highly regulated [107], and malpractice litigation is a perennial concern. Thus, various financial and legal concerns make collaboration between clinicians and amateurs extremely difficult [48].

*LATIN AMERICA.* Historically, Latin American countries share many commonalities in their approach to healthcare [137]. Most Latin American countries provide healthcare to workers via insurance plans. This translates to cases where the working class (typically the lower class) has access to services of variable quality, while the wealthy rely on private services [39]. Recently, however, a series of reforms across Latin America have been implemented to increase the availability of universal healthcare and reduce out-of-pocket expenses [103]. Brazil, for instance, has created the tax-funded "Unified Health System", where health is seen as a right to everyone and an obligation of the government. The Unified Health System, financed through tax revenues, offers Brazilians not only basic healthcare, but also treatments, exams, and medications [93]. Similarly, in 2003, Mexico embarked on a reform to provide healthcare access to all families, especially low-income families, who have been traditionally excluded from the healthcare system [131]. Likewise, through its Social Health Insurance program, Chile offers nearly universal health coverage to its citizens [3]. Costa Rica has also enacted universal healthcare. Compared to the healthcare offered throughout Central America, Costa Rica provides one of the region's best options [139]. However, it is still working to actually provide and sustain universal health coverage in the country.

*INDIA.* The healthcare system in India is stratified with health outcomes that fall along the lines of gender, caste, wealth, education, and geography [13]. India has not implemented a universal healthcare system, although it does have a hybrid system of private and public insurance [60]. The out-of-pocket costs for Indian households are approximately 75% of total healthcare costs [65]. As a comparison, in the case of the U.S. citizen, they only amount to 10% [35]. In fact, healthcare costs for "catastrophic" reasons often lead to extreme poverty for many Indians [65]. Governmental expenditure on healthcare coverage ranges from less than 2% of the GDP [60, 65]. These estimated GDP percentages contrast dramatically with the U.S. rate of 18%. While a much-anticipated universal health insurance is planned for India, it has yet come to fruition [13, 65]. In India, relatively high out-of-pocket costs may drive recipients to consider non-clinical solutions.

*FRANCE.* France has historically been known for its universal healthcare system, "Protection Universelle Maladie" (PUMA) [108]. While the standard of care is very high, depending upon the factors used to assess success, French healthcare rankings vary greatly due to poor economic equity, geographic disparities, and poor administrative efficiency [26]. A Commonwealth fund report [29, 117] ranked French healthcare below many European countries, but not as low as the U.S. Healthcare delivery in France is usually a hybrid or mix of public/private approaches to give people access to healthcare. Many people use the public healthcare provided by the government yet also have additional private insurance, called "Mutuelle", to cover specialized problems not included in public healthcare (including interventions with 3D printed AT). In fact, 90% of French citizens are covered by some form of this hybrid public-private health system [115]. Usually, insurance is offered through employers or bought on the private market. In 2007, France's total expenditure for healthcare reached 11% of the GDP with a minimal out-of-pocket cost due to the public/private hybrid system covering most expenses [26]. There has been limited research on healthcare decision-making by French individuals because "a very small part of household spending" has historically been committed to out-of-pocket expenses [115]. However, the high percentage of GDP for healthcare costs and recent routine economic instabilities have caused real fiscal anxiety for the French government concerning their healthcare system.

SUMMARY OF GLOBAL HEALTHCARE SYSTEMS
- In the U.S., regulation and financial structures may discourage collaboration between clinicians and makers.
- In countries with universal healthcare, cost is less of an issue, but service may be delayed [29].
- Access to healthcare also depends upon many other factors, including a patient's race and gender, as well as their geographic location within the country and the availability of doctors.

# 3 METHODS

Our goal in this paper was to shed light on the different ways a wide selection of stakeholders interact to care for recipients of 3D printed AT. This allowed us to better understand the type of ecosystem setups and collaborations that exist, and the care produced. We can then make socio-technical recommendations to empower more ecosystems to give better care. Our primary approach was participatory action research [37], where we not only studied care ecosystems but also gave back to them by developing well-informed actionable guidelines to improve how people collaborate within the ecosystem.

We spent over two years building a network of trust [47, 134], with stakeholders from two groups: the general e-NABLE organization and e-NABLE related chapters and spin-offs in Latin America. One author, Dr. Jennifer Mankoff, volunteered with e-NABLE, specifically to help facilitate research, including volunteering to review proposals. Furthermore, she helped set up a process for collecting recipient information by developing multilingual questions for recipients to record contact information and to ask for their consent for follow-up research. A second author, Dr. Saiph

Savage, a native Spanish speaker, also volunteered and worked closely for several years with multiple e-NABLE-related chapters and spin-offs in Latin America. Her work includes building 3D printed AT and devising and running interventions to engage newcomers who want to join the chapters. One outcome of this participatory approach was that an early leader of e-NABLE, Dr. Jon Schull, became a co-author of this article.

## 3.1 Interview Protocol

During and after these efforts, we conducted semi-structured interviews from August 2018-August 2019 with 31 stakeholders from seven countries (including some drawn from stakeholder groups with whom we volunteered) to obtain rich qualitative information of the different ways that stakeholders organize care. The interviews helped us obtain information on the different ways that stakeholders operate, as well as recipient satisfaction with their devices. Table 2 provides examples of our interview questions, focused on understanding how ecosystems function. Our appendix provides all the interview questions asked to each stakeholder (built also on questions from related work). Our interviews focused on eliciting information about how ecosystems function, how they drive participation, and the type of participation. We also questioned how the ecosystems were organized, how stakeholders worked together (pain points and high points), as well as the cultural values adopted. Some questions were stakeholder specific (for example, we asked makers to provide details about how they fabricate their devices or clinicians questions about the type of medical care they give recipients). In the case of recipients, we asked for details about the years they have used their devices and whether or not they have had interactions with clinicians or makers regarding these devices afterwards. To facilitate participation, we ensured that our interviews were culturally and linguistically appropriate [90, 123]. To this purpose, we held interviews in Spanish (three authors are native Spanish speakers) or English when appropriate. Our team did not include a Hindi speaker. However, Indian participants were all fluent in English. We also first checked with members from each population on the appropriateness of our questions.

| | |
|---|---|
| Ecosystem Interactions | *Could you describe your interactions with makers? with recipients? with doctors/nurses/therapists? What are some of the challenges you encounter when connecting with them? How do you interact with them? (email, phone, video conference, etc)? What type of things do you do with them? Describe the interactions that you considered to be useful. Describe the interactions that you wished had been different.* |
| Ecosystem Culture | *What are some of the values around 3D printed AT that you believe resonate the most with makers? with recipients? and doctors/nurses/therapists? Who are some of their role models? Why?* |
| Ecosystem Work Dynamics | *Is there a structured process for working together with makers, doctors/nurses/therapists, and recipients? (Please Describe.) Give me an example of a time when you worked well as a team with makers, doctors/nurses/therapists, and recipients. Tell me about a time when you found challenges collaborating as a team with recipients, other makers, and doctors/nurses/therapists. How do you stay organized when working with makers, recipients, and doctors/nurses/therapists?* |

Table 2: Example of some of our ecosystem focused interview questions

## 3.2 Recruitment and Participants

Participant recruitment was done separately from our direct engagement with e-NABLE; it was primarily led by students unknown to the stakeholders being interviewed. This helped to ensure that interviewees saw their participation in the study as voluntary. Recruitment began with emails to individuals who had agreed to be contacted (such as e-NABLE recipients who had consented to follow-up research) or individuals who were introduced to the authors by members of the organizations with whom we worked. We also searched for participants using social media and news reports, as well as a snowball sampling to invite more individuals.

Details about our 31 participants can be found in Table 3. Participants included 16 recipients (R1-R16) - the majority from the U.S. and Mexico - but we also had one from Chile, nine makers (M1-M9) from the U.S., Brazil, France and India, and six clinicians (C1-C6) from Mexico, India, U.S., and Costa Rica. While the recipients in our study primarily received devices that function like "hands", we consider our study to be about upper extremities because makers and clinicians also work with recipients to address other customized needs. The recipients themselves also learned broadly about 3D printed AT.

In order to maximize the variety of ecosystems we could study, we emphasized recruitment from multiple structural contexts, including industry/volunteer/government and groups operating in seven different countries (See Fig. 2). Another important recruiting goal was to include recipients (those who had received and used 3D printed assistive technology), makers (who made 3D printed AT), and clinicians (who provided any type of medical assistance or therapy assistance for the operation of the devices) - ideally all three in the same ecosystem. As shown in Figure 2, a total of 10 care ecosystems were represented through the recruited participants. We succeeded in including at least two stakeholders in over half of the ten ecosystems, and all three stakeholders in two of the ten. Recipients are represented in six different ecosystems (three different countries). Similarly, makers are included in six ecosystems, including four countries, and clinicians are included in six ecosystems, including four different countries. All stakeholders are represented in at least one corporate and multiple NGO contexts. Table 3 provides more details on the participants. The makers in our study volunteered for e-NABLE or worked at companies, governments, or hospitals. The clinicians also worked in hospitals, governments, and university settings; several were associated with e-NABLE in some way. Of note, some clinicians also self-identified as makers. Some recipients manufactured their hands themselves and self-identified as makers. Most recipients obtained their hands from e-NABLE chapters, hospitals, or companies.

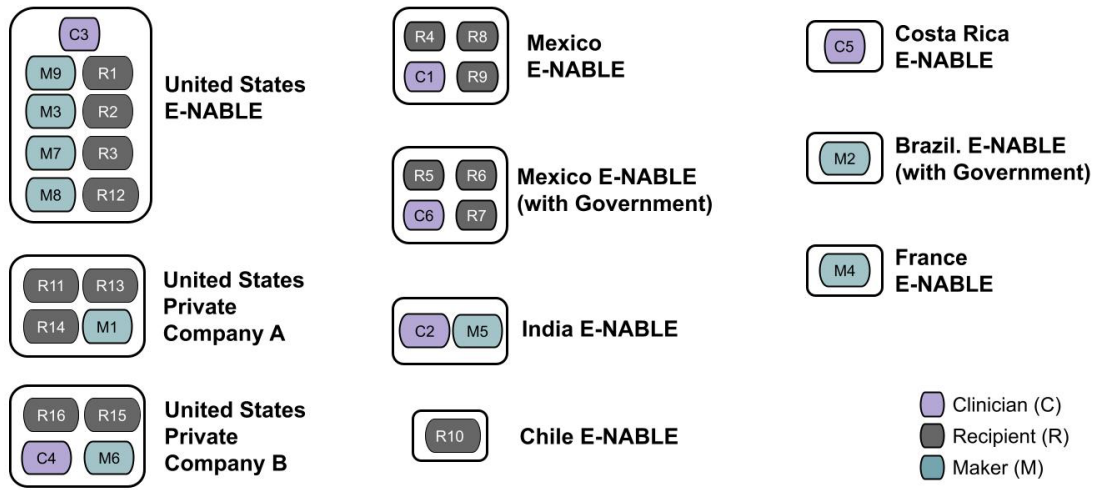

**Fig 2:** Overview of the care ecosystems of our participants. Title for each grouping indicates the type of entity leadership that it is primarily associated with. Participant labels indicate whether the participant is a (C)linician, (R)ecipient or (M)aker.

## 3.3 Data Analysis

Interviews were transcribed and aggregated with notes and memos from the study. We used open coding to extract initial concepts from the data [96], using a combination of bottom-up theme extraction and top-down themes derived from related work [48, 73, 100]. As a group, three of the authors first independently coded the data bottom-up and together developed a set of eighteen axial codes that were applied top-down to the entire interview transcripts. Our analysis showed strong inter-coder agreement on the axial codes (Cohen's Kappa coefficient (k) = 0.765). Disagreements were discussed during the writing and synthesis process. From the eighteen top down axial codes, we developed five themes that organized the main insights from our semi-structured interview transcripts. We have summarized each theme:

**Collaboration approaches:** this theme includes the collaboration approaches that ecosystems have adopted and the challenges faced, such as: (a) limited maker collaborations with clinicians, and (b) how interactions between the members of the ecosystem impact care.

**Structural support:** this theme includes the resources that people have adopted in their ecosystem to provide care.

**Identity:** this theme examines how the identity people adopt relates to care experiences with 3D printed assistive devices and shapes dynamics within the ecosystem; e.g., recipients who also identified as makers, as well as clinicians who self-identified as makers.

**User perspectives and experiences:** this theme examines the perspectives and experiences that recipients had with their devices and how other actors dealt with recipients' concerns and feedback.

**Maker culture:** this theme examines how the maker culture influences participants' work dynamics.

| | | | | |
|---|---|---|---|---|
| R1 | Recipient & Maker | U.S. | Hand from e-NABLE | Male |
| R2 | Recipient | U.S. | Hand from e-NABLE | Female |
| R3 | Recipient & Maker | U.S. | Hand from e-NABLE | Non-Binary |
| R4 | Recipient | Mexico | Hand from e-NABLE | Female |
| R5 | Recipient | Mexico | Hand from hospital | Female |
| R6 | Recipient | Mexico | Hand from hospital | Male |
| R7 | Recipient | Mexico | Hand from hospital | Male |
| R8 | Recipient | Mexico | Hand from e-NABLE | Female |
| R9 | Recipient & Maker | Mexico | Hand from e-NABLE | Female |
| R10 | Recipient | Chile | Hand from hospital | Female |
| R11 | Recipient | U.S. | Hand from company | Male |
| R12 | Recipient | U.S. | Hand from e-NABLE | Female |
| R13 | Recipient | U.S. | Hand from company | Male |
| R14 | Recipient | U.S. | Hand from company | Male |
| R15 | Recipient | U.S. | Hand from company | Male |
| R16 | Recipient | U.S. | Hand from company | Male |
| M1 | Maker | U.S. | Works at company | Male |
| M2 | Maker | Brazil | Collaborates with e-NABLE, government, and hospitals | Male |
| M3 | Maker | U.S. | Collaborates with e-NABLE | Male |
| M4 | Maker | France | Collaborates with e-NABLE | Male |
| M5 | Maker | India | Collaborates with e-NABLE and government | Male |
| M6 | Maker | U.S. | Works at company | Female |
| M7 | Maker | U.S. | Collaborates with e-NABLE | Female |
| M8 | Maker | U.S. | Collaborates with e-NABLE | Male |
| M9 | Maker | U.S. | Collaborates with e-NABLE | Female |
| C1 | Clinician & Maker (Medical Maker) | Mexico | Collaborates with a hospital and e-NABLE | Male |
| C2 | Clinician & Maker (Medical Maker) | India | Collaborates with a hospital and e-NABLE | Female |
| C3 | Clinician | U.S. | Retired occupational therapist volunteering at e-NABLE | Female |
| C4 | Clinician & Maker (Medical Maker) | U.S. | Works at hospital | Female |
| C5 | Clinician | Costa Rica | Government physician; collaborates with e-NABLE; and university | Female |
| C6 | Clinician | Mexico | Government physician; collaborates with e-NABLE; and university | Male |

**Table 3:** Overview of the characteristics of our participants

## 4 RESULTS

In this section, we first use a thematic analysis of our interviews to characterize the different ways that care ecosystems collaborate and work. (We connect here with the theme of "Collaboration Approaches"). Next, we focus on identifying how the different types of collaborations impacted care (connecting with the theme of "User Perspectives and Experiences"), uncovering exactly how some ecosystems managed to provide better care than others. (Here, we connect with the themes of "Identity" and "Collaboration Approaches"). We finish by presenting concrete features that ecosystems have in place to provide quality care for recipients. (In this part, we connect with the themes of "Structural Support", "Collaboration Approaches", and "Maker Culture").

### 4.1 The Importance of Ecosystem Structure in Supporting Follow-up

Prior work has established the risks of abandonment with upper-limb prosthetics [20], some suggesting the importance of "follow-up" care in the volunteering context (e.g., [101]). Our interviews highlighted the range of challenges that follow-up must address, such as fixing broken assistive devices, adjusting the device to better fit the recipient, or even providing recipients with physical therapy that helps them to better use their devices. Ecosystem processes strongly impacted abandonment. Some participants had lengthy and detailed processes for providing follow-up. In contrast, for others, it was completely missing from their ecosystem, and they did not seem to know how to implement these care activities. Participants M1, M2, M4, M5, M6, R4, R5, R6, R7, R8, R9, R10, R11, R13, R14, R15, R16, C1, C2, C4, C5, and C6 reported being in ecosystems, where follow-up was provided to recipients, while M3, M7, M8, M9, C3, R1, R2, R3, and R12 reported being in ecosystems where follow-up was rare. In situations where follow-up was lacking, the recipient experience was negatively impacted. For example, R1, R2, and R12, all came from ecosystems with limited maker-clinician interactions and no follow-up. They all reported that the original 3D printed assistive device they had received

Savage et al.

from their ecosystem was unusable, likely because the lack of follow-up meant that their devices were never improved or better tailored to them. For example:

> "...it's been a year since the first inception of receiving the hand [3D printed AT] and it's kind of been just sitting in her box not doing anything ... [the device] doesn't feel good, doesn't work well...", R2, USA.

The lack of fit is a well-known reason for abandonment [20]. Compare this to R8, a recipient who received follow-up from a clinician, who taught her best practices for using her AT:

> "Once the device is ready [i.e., once makers have finished fabricating the 3D printed assistive device], they call us back. We pick it [the device] up and set up a schedule for therapies [with clinicians] so we can learn how to use the device...", R8, Mexico.

At the time of our study, R8 had been using her device for over two years. Similar results were found for recipients R4, R5, R6, R7, R8, R9, R10, R11, R13, R14, R15, and R16 – all part of ecosystems, where clinicians worked closely with makers to provide follow-up that included therapeutic support. All reported using their devices for over two years at the time of the study.

### 4.1.1 MULTI-STAKEHOLDER COLLABORATION FOSTERS FOLLOW-UP.

In ecosystems that provide follow-up, all had collaborations between clinicians, makers, and recipients, with clinicians often leading the effort. For example, M2 describes how his ecosystem involved clinicians from the start, and they were the key decision makers:

> "... the first step is that we [his team of makers] ask [recipients] two things. First: do you have a doctor? If he has a doctor, we schedule a meeting with this doctor ... and we ask this doctor, who knows about the case and the patient, to evaluate the devices and based on that to say what is the device we need to create for the kid [recipient]. So, we [the makers] do not choose, also the recipient does not choose. The doctor is the one who will tell us what needs to be printed ...If the kid doesn't have a doctor ...we try to find a doctor near the city ... and we approach this doctor ... and ask him to help this person with his evaluation. So, the doctor is the first step...", M2, Brazil.

Having a setup where clinicians interact closely with makers likely helped cover makers' knowledge gaps in medical topics, while helping to deliver better devices and interactions for recipients (including follow-up). In ecosystems where clinicians worked closely with makers and recipients, it was easier for all stakeholders to share their specialized knowledge to provide follow-up and ensure quality experiences for recipients. Participants M1, M2, M4, M5, M6, R4, R5, R6, R7, R8, R9, R10, R11, R13, R14, R15, R16, C1, C2, C4, C5, and C6 expressed that within their ecosystems, all three stakeholders had sustained interactions with each other, helping to provide follow-up and appropriate AT design for recipients. The following maker from e-NABLE, M5, shared how in his ecosystem, makers, clinicians, and recipients worked together to deliver quality care and follow-up:

> "... the engineer [maker] is responsible for the design work and the modifications that need to be done on the printers and the design software. But the prosthetic technician [medical maker] gives him [the maker] ideas on how he can edit the designs [. . . ] we have another occupational therapist [clinician] who lives in New Delhi, who guides the doctor [clinician] on all the training [the 'training' here refers to the follow-up training that is given to the recipient to help the recipient learn how to use the 3D printed device], on like what kind of trainings he [the doctor] should be giving [. . . ] We also ask the beneficiaries [recipients] for ideas just like how we asked [recipient] what works better...", M5, India.

Another benefit we observed of maker/clinician collaborations was having follow-up that quantified how assistive devices were used by their recipients. For instance, M6, a maker who developed 3D printed AT within a private company, worked closely with a clinician who developed a follow-up protocol for measuring the way that recipients used their assistive devices. M6 explained how this follow-up protocol used evidence-based medicine practices to decide the best way to care for individual patients [114]:

> "... so, I am friends with [name changed], a hacker who is at the School of Medicine. One of his goals for the last years has been improving evidence capture ...He's also developed one of the more recent novel upper extremity outcome [assessment] protocols. To me it's important to be using upper extremity outcome assessments for my designs... ", M6, USA.

Such data can help define strategic design modifications and identify the ideal medical care for recipients.

ACM Transactions on Accessible Computing, Vol. 15, No. 4, Article 31. Publication date: October 2022.

### 4.1.2 LACK OF COLLABORATIONS WITH CLINICIANS MAY INCREASE RISK.

Clinicians who came from ecosystems, where they had limited interactions with makers, shared concerns regarding makers' medical knowledge gaps. C3 came from an ecosystem, where clinicians and makers rarely collaborated. As a retired hand therapist, she observed the following:

> "The most glaring one [issue] that I saw was on the prosthetic [3D printed assistive device] that they handed out. It [the 3D printed AT] was very poorly fitted on the residual stump [limb of the recipient] and in order to correct this the [leader in the ecosystem] tried to remediate this by tightening the straps really really tight on the gauntlet [a part of the 3D printed assistive device]. But this would not be suitable because to [a clinician] it would be causing pain for the recipient and it could also lead to nerve compression if it was left on too long.", C3, USA.

Fit was a major concern for C3. She believed that the makers in her ecosystem did NOT fully account for the medical needs of recipients in their designs, nor in their interactions with recipients.

## 4.2 Maker Acculturement: Adoption and Spread

Our interviews exposed how several recipients and clinicians also self-identified as makers. These crossover participants had the advantage of directly addressing some of the challenges associated with providing/using the devices themselves.

### 4.2.1 FROM CLINICIAN TO MEDICAL MAKER.

In the ecosystems where clinicians actively collaborated with makers, the clinicians tended to take a more active role in the fabrication process, with some of these clinicians transitioning to the role of being "medical makers" (i.e., clinicians who are also makers [74]):

> "...we have a prosthetic technician. So, he [a medical maker] assesses the hands and he modifies the hands [3D printed AT]. He tells us what tools to use and when we have a challenging beneficiary [a recipient with a medical condition which requires a new or adapted AT design], he gives us ideas on how we can modify the design ...", C2, India.

Such "medical makers" had the advantage of more easily providing follow-up, which could include customizing or fixing devices to better address the medical needs of the recipients. Another advantage was that these clinicians were able to more easily collaborate with makers because they had access to a shared design language.

### 4.2.2 RECIPIENT MAKERS: A GROWTH MINDSET.

Similar to clinicians, the recipients of 3D printed AT can be exposed to makers and their cultures [1]. Almost all our recipients (R1, R2, R3, R4, R7, R8, R9, R10, R11, R12, R13, R14, R15, R16) expressed that they identified on some level with the values of the maker's culture, such as "doing things by oneself", "sharing knowledge", and having a "growth mindset" (referring to the belief that a person's capacities and talents can be improved over time) [66]. Note that "identifying with the maker culture" does not mean that these recipients necessarily participated in the fabrication and direct modification of the devices. Identifying with the maker culture and adopting the mindset of finding value in its related growth helped the recipients to view their devices as something that was not yet finished. They viewed their devices as a "work-in-progress" that was not perfect but could be improved:

> "[I] couldn't hold a spoon [with the device], could grip, but not eat with it. Went to the dollar store. Tried Velcro, PVC pipe. I would put the Velcro on the hand and on the end of that PVC pipe. And I'd put the fork inside that PVC, and that worked quite well at first but it just wasn't strong enough. Then got magnets. Oh and the magnets worked so well. Got bar magnets and good glue and I glued the magnet to the palm of the hand [3D printed AT]", R1, USA.

Fig. 3 illustrates an example of the maker activities that R1 conducted to modify his 3D printed assistive hand so he could ensure being able to eat with the device.

> "... the device [3D printed assistive device] is a great way to induce: a growth mindset . . .I saw the device [3D printed assistive device] as a prototype, something that needed to be changed and edited...", R3, USA.

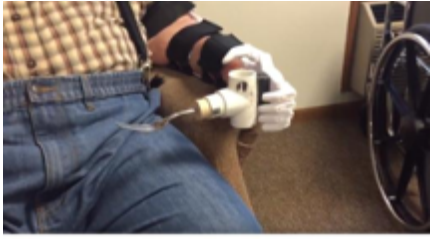

Figure 3: Example of device adaptation made by one of our recipients, who also identified as a maker

This notion of identifying with the growth mindset and hence seeing their devices as something to be improved over time lead some recipients to decide to directly address the challenges of follow-up by taking part in maker activities. These activities were a way through which recipients could take direct action to fix and improve their devices themselves. Recipients R3, R7, R10, R13, R14, R15, and R16 also shared that identifying with the maker culture helped them view their disability and the related care they sought, as something that could be improved over time. The growth mindset of the maker culture helped the recipients view the activities they could do with their limbs as something to be improved over time. Identifying with the growth mindset also helped these recipients value their therapies more, as shared by R10:

> "...I am doing my extensions [therapy exercises] all the time. While I'm watching TV, chatting with family or just relaxing I'm always doing my extensions [therapy exercises]. But all the time I've spent [on the therapy exercises] has been advantageous. I've improved a lot . . .I am never going to stop [doing the therapy exercises] till I get it [3D printed assistive device] to work like the hand I had.. . . ", R10, Chile.

### 4.2.3 A MAKER FAMILY/COMMUNITY.

The recipients' membership in a care network extended also to their families. For example, R1, R3, and R9 all reported that their families engaged in making with them:

> "...It is usually the women, like my mom or other moms who become makers to help their kids.. . . ", R9, Mexico.

The recipients who reported having their families involved in the making process were primarily those recipients who self-identified as makers. (Overall, these recipients not only adopted the maker culture, but they also actively participated in the design and fabrication of 3D printed devices). Having family members who identified as makers helped them invest more time in fixing and improving their devices. R3, an e-NABLE recipient, built their 3D printed AT with their father. Although their 3D printed device was not perfect, the fact that a parent had helped build the device motivated them to want to invest more time in improving their device as it had become an activity they could do together:

> "...My dad and I built the hand [the 3D printed assistive device] together. The first version [of the device] hurt a lot. It was a prototype, and there was always something that needed to be changed and edited. But because my dad and I made it together, it was a project we wanted to finish . . . Part of it was that it was fun to make changes, and make them together.. . . ", R3, USA.

Within these bonding activities, some recipients also reported benefiting from their connections with other recipients. R1, R3, R7, R9-R11, R13-R16 described that they practiced "knowledge sharing" to learn from each other about new care techniques. Some of these care techniques included "new rehabilitation exercises", "ways to psychologically deal with limb loss", "how to do by themselves certain activities", "which 3D printed material was best for specific devices", or even "how to get better medical insurance that would cover the costs associated with their 3D printed devices":

> "...We [group of recipients] show up and just share how we do things with our hands [3D printed assistive devices]. One same activity is done differently by each person . . .It's about learning together new ways to live, new ways to live with our new hands [3D printed AT].. . . ", R7, Mexico.

> "... part of my goal [on an online forum from his ecosystem] is having open discussions about what's working, for each of us, and what's not working. How did you get the insurance company to pay [i.e., cover the costs of a private company fabricating the device and providing follow-up]?...", R16, USA.

These knowledge-sharing activities took place primarily within online forums or physical meetups (e.g., convention centers or casual in-person reunions). Recipients shared that they were invited through their ecosystems to join these social gatherings. The social gatherings also helped for emotional support:

> "I see people...[recipients in the social gatherings of his ecosystem]...who say: "I have to go to surgery tomorrow!" and I'm like, Just take what I'm saying and put in your heart and keep it there, you're going to be okay and you will adapt and you're going to be a better person. . . ", R14, USA.

## 4.3 Ecosystem Services

To better understand the role of the overall ecosystem in fostering success, we explored the ecosystem services that may have helped foster follow-up.

### 4.3.1 PREPARATORY WORK.

One unusual aspect of some ecosystems was the degree of preparatory work fostered. Unlike U.S. maker communities that emphasize rapid production [48], some ecosystems and recipients studied valued a slower approach to designing successful AT. For example, R5, a recipient who lost her hand in an accident, shared how she had one-on-one follow-up sessions to plan how her device would be modified to better fit her needs:

> ". . . These sessions [one-on-one sessions with her medical maker] have helped me to start to craft with my doctor my ideal device . . .My ideal device is one that will help me to re-establish the image I had about myself before the accident and with the same functionalities that I had before", R5, Mexico.

The ability to address aesthetic goals and custom needs is one advantage of 3D printed AT since it leverages a volunteer maker network willing and able to take the time to customize the look of the designs, as well as prioritize different functionalities. These consequences help the device to be used long-term, something that may be harder to address in more traditional clinical settings [111]. Furthermore, engaging recipients (and makers) in such an iteration provides an opportunity for learning maker values as well as human-centered design skills. Together, this fosters identification with and adoption of those values and skills.

### 4.3.2 ESTABLISHMENT OF FORMAL COLLABORATIONS.

We found that ecosystems with sustained follow-up have established formal collaborations among the different stakeholders. It is known that having formal collaborations can help avoid uncertainties between the different stakeholders long term [55]. Therefore, the integration of formal agreements likely facilitates the sustained participation of all stakeholders and helps to ensure follow-up with input from makers, recipients, and clinicians. Our interviewees discussed how these formal agreements took the form of insurance policies (R11, R13, R14, R15, R16, M1, M6, and C4) or officially signed documents (R4, R5, R6, R7, R8, R9, R10, M2, M4, M5, C1, C2, C5, and C6). C6 shared how his ecosystem had an official agreement between local government hospitals and universities (which had e-NABLE chapters). The agreement detailed how the different stakeholders would interact to produce 3D printed assistive devices and provide quality experiences to recipients:

> ". . .we had to go to the doctor who is the director of [government clinics in a region of Mexico] . . . [This doctor] signed off on the project, and he was going to be responsible for overseeing that the devices did not have any side effects on patients . . .we also had to get professors [from universities with e-NABLE chapters] to sign off that they would help with the 3D printing.. . . ", C6, Mexico.

Similarly, M2, who came from an ecosystem with follow-up and sustained participation from all stakeholders, reported that in his ecosystem, the makers did not produce devices if they did not first receive a signed and certified letter from clinicians, detailing their agreement to help recipients, as well as showcase proof of the required abilities and knowledge to support recipients:

> "Here, we [makers] only donate the device if the physiotherapist [clinician] sends us the prosthetic prescription with signature and professional registration number [indicating the clinician is trained] . . . [clinicians] register as volunteers and are trained to know how to prescribe the devices.. . . ", M2, Brazil.

These formal collaborations were missing from the ecosystems that did not provide follow-up. M3, M7, M8, M9, C3, C4, R1, R2, R3, and R12 described how the different stakeholders only interacted "casually" in their ecosystems. M8 shared how he only sometimes "casually" connected with clinicians through forums. But no formal collaborations had been established:

> "In terms of medical professionals [clinicians], I haven't worked directly with a lot of medical professionals. There have been a few prosthetists [clinicians] that have worked with us [his ecosystem] just through the online community. And my interactions with them have been mostly just through those online, you know communications...", M8, USA.

### 4.3.3 SECURING FUNDING FOR COLLABORATIONS.

Another strategy found in the ecosystems with follow-up was that some "secured funding" to ensure that certain collaborations took place. The funding was to cover the participation costs of clinicians (R11, R13, R14, R15, R16, M1, M6, C4, C2, C5, M5, and M4), makers (R11, R13, R14, R15, R16, M1, M6, C2, C4, and C5), or even recipients (M5, C1, C2, C5, and C6). C2 shared how in her ecosystem, they had raised funds to cover the participation costs of all stakeholders to facilitate their participation. This was especially important in the case of recipients who often lacked the resources to travel to their follow-up:

> "...So we raise funds and no one is volunteering [for the follow-up], we employed them all [...] some of them [recipients] didn't come initially [to the follow-up] because they didn't have the money. [Recipients are] really, really poor so what [the ecosystem] started to do is that...we pay their transport and give [recipients] a meal so that they come to us. So that motivates [recipients] to come...", C2, India.

Funding to support participation was missing from those ecosystems that lacked follow-up; however, not all of the ecosystems with follow-up implemented monetary payments. This was particularly true for our interviewees from Brazil, who were able to secure resources to cover participation costs. But these resources were not directly translated into monetary compensations, as they felt it could demotivate participation. Therefore, this ecosystem made a point of not giving anyone monetary payments:

> "...One of the basis of the work is that all volunteers, even the doctors, need to do the work free of charge [...] the best way to get people to stop doing what they like is paying them to do it. Imagine you like to paint and you do it for free, because you like it. So, I start paying you one dollar for each painting; with time, you will think that your work is not being valued and you will want to receive more for it. So when we come to volunteer work, the idea is to engage the volunteer in other ways that don't involve money....", M2, Brazil.

It was interesting to observe that this particular ecosystem avoided, in general, any type of monetary transaction. The resources they donated were supplies and materials to fabricate the assistive devices:

> "...We avoid using money in every part of the process [...] we also do not accept donations in money. Only donations of raw material and equipment used to create the devices....", M2, Brazil.

**SUMMARY OF RESULTS:** We present some of the key findings from our results:
- Providing follow-up to recipients was a key challenge that ecosystems faced (with some ecosystems never able to provide sustained follow-up).
- Ecosystems with multi-stakeholder collaborations tended to produce better care, in terms of device adoption and reported recipient experiences.
- The exposure of the maker culture to recipients and clinicians had positive care outcomes (in terms of recipient satisfaction and device usage).
- To involve multiple stakeholders, ecosystems adopted different strategies for motivating participation, such as creating formal collaborations around stakeholder involvement and securing funding to cover the participation costs, when needed.

## 5 DISCUSSION

In this paper, we studied the different ways that care ecosystems organize to provide care to recipients. During this discussion, we focus on key features that characterize how these care ecosystems in 3D printed AT function. Overall, our study highlighted how care ecosystems with organizations like e-NABLE, have helped to address important gaps in institutional healthcare services. The care ecosystems were especially useful for those who experience the gaps (usually individuals from underserved populations, who tended to not be found in institutional healthcare systems, nor most academic settings, e.g., people from regions in India without the economic resources to even be able to attend follow up).

The care ecosystems can become a social safety net for a wider range of individuals. Whether society recognizes and supports the value of this safety net will depend in part on sustained interaction and accommodation among institutional and non-institutional stakeholders. This study suggests that an ecosystemic perspective will help ensure that people get the best possible care when institutional solutions are available and when systemic failures occur.

## 5.1 Integrating Formal Collaboration in the Maker Culture

Formal collaborations are common in clinical settings; it is usually expected that clinicians will work with finished and well-tested care products [18, 110]. Within formal settings, usability, safety, and the functionality of care products is typically preferred over novelty [34, 62]. Makers, on the other hand, usually value rapid and novel device production over more limited, but safer, designs [50]. This attitude can cause serious conflict between clinicians and makers.

For example, one study reported about "violent language" and emerging threats when a maker community was questioned about the safety of the personal protective equipment they were fabricating during the onset of the COVID-19 pandemic [50]. Conflict may also arise because the maker culture pushes heavily for collaborations to be part of a "casual affair," where work and leisure are intermixed and not easily distinguishable [82]. However, our interviews highlighted how the ecosystems most effective at providing care had formal collaborations in place with expectations about training, deliverables, and non-making activities.

Another difficulty that emerges from formal collaborations centers around the assessment of outcomes (e.g., quality metrics [84]). It is expected that low-quality work will be rejected. However, within the maker culture, this may clash with the idea that making should be an activity that people do out of their own passions and interests. Suddenly rejecting work done "for fun" and in a volunteer setting can have a challenging or discouraging impact.

These formal collaborations facilitate sustained, multi-stakeholder participation as stakeholders have to first take the time to understand the needs and context of the other actors in order to create an agreement that would be reasonable for everyone to take part in. Notice that establishing formal collaborations consequently helps identify challenges, including those of certain stakeholders, which then facilitates creating new mechanisms to address those challenges (e.g., designing ways to cover the recipients' costs, who originally could not afford to attend their follow-up). We argue that care ecosystems might consider establishing formal collaborations with all stakeholders. Formal contracts took different forms across the ecosystems we studied. In short, there is no one correct solution.

Ecosystems need to explore the type of instantiation of formal agreements that makes the most sense, given their cultural and local settings. For instance, in the U.S., clinicians may be putting themselves at risk if they collaborate closely with amateur makers [48]. Similarly, in India, people might be more accustomed to directly paying for their own healthcare [13, 60, 65]. Consequently, they might be more open to having a formal contract, where they directly pay clinicians and makers to help in the design and fitting of their 3D printed AT.

On the other hand, when thinking about integrating clinical structures within a creative amateur movement, such as e-NABLE, it can help to think about what may be lost with greater formality. Recent work focused on COVID found that some makers choose to go their own way when confronted with safety guidelines [50]. Future work could explore the values held by stakeholders and how safety regulations can support multiple value systems, as well as the possible tradeoffs/benefits of adopting more or less formal collaborations.

## 5.2 The Value of Compensated Participation

A portion of the maker culture argues that makers should not be paid monetarily for their labor [56] as it is expected that their true reward is building things for which they are personally passionate (i.e., they are expected to be intrinsically motivated) [71, 77, 109]. However, not paying people for their participation can also limit who takes part [32, 70], privileging those with sufficient resources. Our interviews highlight that ecosystems with follow-up cared more about ensuring the participation of more stakeholders, especially low-income ones, via whatever means made sense, e.g., giving monetary compensation to cover participants' transportation costs or meals.

Our interviewees came from diverse sectors, e.g., government, NGOs, academia, and industry (see Table 3). Therefore, it is unreasonable to think that one type of compensation will always function to motivate everyone to participate. In India, monetary compensation for recipients might be especially important given that the country's healthcare system does not usually cover patients' costs [60], leading a large number of Indians to "go broke" when trying to cover their medical needs [65]. But monetary compensation might not necessarily work well when trying to involve academics in the ecosystem. However, offering research opportunities enhances professional and social status and could drive their participation [4, 97, 98]. Similarly, as some interviewees mentioned, people's intrinsic motivation could be negatively affected when they suddenly start receiving monetary payments [124, 130]. Ecosystems will benefit from hybrid approaches for compensating and ensuring the participation of all stakeholders.

## 5.3 Tools for Sustained Follow-up and Multiple Stakeholder Participation

Our findings emphasize the importance of including all stakeholders in the process of providing care. This is fitting given that AT provision and follow-up requires mixed expertise [126]; clinicians are needed to ensure that the follow-up does not cause medical harm to recipients and foresee that the activity helps recipients in their rehabilitation [15, 48, 75, 83]. Recipients also need to be included to provide direct feedback and suggestions on the therapy and device modifications that take place during the follow-up [92, 100]. Similarly, makers' expertise is needed to facilitate device modifications [1, 135].

In other words, sustained participation of all stakeholders is important to ensure the production of quality AT and implement follow-up [53, 57]. Toolkits for fabricating 3D printed AT should, therefore, aim to integrate guidelines and computational mechanisms to motivate stakeholders to engage with each other. Having more continuous interactions among stakeholders will facilitate follow-up and improve outcomes. Existing toolkits (such as e-NABLE webcentral[3], or Latin American efforts[4]) might be a basis for this, but we argue that they should also integrate mechanisms to facilitate collaboration and the discussion of factors that impact all aspects of a device's lifecycle, including safety, fit, design aesthetics, and follow-up.

The best approach is yet to be determined, but possibilities include: a forum, a ticketing system for recipients to report concerns, digital badges, a mini-course on how to include and work with other stakeholders in the fabrication of 3D printed AT, and even gamification. (It is important to note that some of these solutions, especially the ticketing system, and badges, have recently been implemented by e-NABLE as a result of this research.) Such a tool could also facilitate compensation in multiple forms – tracking of donated time and materials, the transfer of funds to recipients to encourage them to return for follow-up services, and so on. We also believe there is value in designing a recipient role into such a toolkit. This could guide recipients not only on how to best use their new devices, but also how to connect with the other stakeholders in their care ecosystem, as well as other recipients for support and idea sharing. Such a toolkit could provide guidance on the type of feedback recipients should provide to clinicians and makers to help them provide quality care in the best way possible. There is likely value in integrating mechanisms in these toolkits to help recipients learn about fabrication. (In our interviews, we saw how recipients who self-identified as makers used their devices longer term and were overall more satisfied).

## 5.4 Storehouse for Cross-Ecosystem Collaborations

Given the variety of ecosystem structures observed, there is value in helping stakeholders learn what other ecosystems are doing in combining making with medical practices to improve the recipient experience. The problem is that state-of-the-art tools [23, 27, 113] currently do not facilitate collaboration across ecosystems, let alone the collaboration of clinicians, makers, and recipients across countries. For example, the NIH 3D Print Exchange [27] is an online space that allows makers to find, create, or share 3D-printable models that are scientifically accurate or medically applicable.

While the NIH 3D Print Exchange is an innovative space to engage makers in biomedical science, it does not support the device lifecycle past design and a safety review. It also does not support recipient involvement. Additionally, the site's overall design assumes that people are participating as individuals instead of an ecosystem. This means that the platform does not necessarily facilitate cross ecosystem collaborations, whereby ecosystems could help each other meet important challenges. A more comprehensive storehouse might be expanded with meta-data to support licensing (or sharing agreements around distribution and ecosystem functioning), as well as other meta-data related to the collaborative care practices surrounding design and safety. Quality information includes design variants, known concerns, and parameter limits (such as minimum and maximum print sizes), along with follow-up information for clinicians on the type of therapies to implement to help recipients adopt a particular AT design while minimizing harm. The meta-data could also provide information about the common problems recipients have encountered with their devices and ways in which makers can address these issues. This meta-data could include advice for recipients on where and how to fix their devices or where to receive therapy to help in adapting to them. Notice that this type of meta-data would be aligned with some of the design recommendations of prior work [63], which argued the importance of providing a transparent information flow between the different stakeholders for successful collaborations. Note that as a result of this research, e-NABLE has recently started to implement several of these meta-data recommendations. (See the new e-NABLE device catalog[5].)

---

[3] https://www.enablewebcentral.com/
[4] https://limbs.earth/

**Limitations** Our research is a snapshot in time of the stakeholders involved in 3D printed assistive technology for upper extremities. As time passes, the ecosystems will continue to evolve. Hence our study only reflects those interviewed at a specific time. In addition, like any study dependent on volunteer participants, our participants' opinions may be biased by self-selection. We were not able to obtain an equal number of stakeholders from each ecosystem, given the difficulty in connecting and finding them. However, given that previous research has not been able to understand how these ecosystems have functioned outside the U.S. and Canada, we consider our study to be a step forward in understanding more about the ecosystems emerging around 3D printed AT for upper extremities. Future research might include more voices from other countries.

# 6 CONCLUSION

HCI has a growing interest in studying the complex dynamics in care ecosystems, then using this understanding to design tools that will enable better experiences for people with disabilities [49, 51, 73, 88, 100]. We broadened the HCI research by studying the care ecosystems surfacing around recipients and their 3D printed assistive devices. To widen our knowledge of these care ecosystems, we included stakeholders from seven different countries (Brazil, Chile, Costa Rica, France, India, Mexico and the US). We do not claim to have studied all possible care ecosystems. We simply bring a more diverse view than the one-country studied by prior work. This has allowed us to uncover new findings about ecosystem services and properties that facilitate success. Through our research, we identified the ecosystems that provide the most effective care (in terms of recipient satisfaction and device usage) as those ecosystems that created formal collaborations around stakeholder involvement, provided compensations to cover participation costs when needed, and had clinical involvement in the ecosystem. We discussed how ecosystem goals can conflict with the maker culture driving 3D printed AT. However, we presented pathways to address these conflicts and challenges. Our paper ends by discussing design implications from our findings to create tools that support multi-stakeholder care ecosystems.

**Acknowledgments.** Special thanks to all the anonymous reviewers who helped us to strengthen the paper. A huge thanks to all of our participants, especially he global e-NABLE community who have been a key ally for conducting this research. This work was partially supported by NSF grant FW-HTF-19541.

---

[5] https://hub.e-nable.org/p/devices